# A Comprehensive Simulation Study of a Liquid-Xe Detector for Contraband Detection

I. Israelashvili[1,4], M. Cortesi[2], D. Vartsky[1], L. Arazi[1], D. Bar[3], E. N. Caspi[4] and A. Breskin[1]

[1]*The Weizmann Institute of Science, Rehovot 76100, Israel*
[2] *National Superconducting Cyclotron Laboratory (NSCL), East Lansing 48823 (MI), USA*
[3]*Soreq Nuclear Research Center (SOREQ NRC),Yavne 81800, Israel*
[4]*Nuclear Research Center of Negev (NRCN), Beer-Sheva 9001, Israel*
Corresponding Author: israelashvili@gmail.com

**ABSTRACT:** Recently, a new detector concept, for combined imaging and spectroscopy of fast-neutrons and gamma was presented. It encompasses a liquid-xenon (LXe) converter-scintillator coupled to a UV-sensitive gaseous Thick Gas Electron Multiplier (THGEM)-based imaging photomultiplier (GPM).

In this work we present and discuss the results of a systematic computer-simulation study aiming at optimizing the type and performance of LXe converter. We have evaluated the detector spectral response, detection efficiency and spatial resolution for gamma-rays and neutrons in the energy range of 2-15 MeV for 50 mm thick converters consisting of plain LXe volume and LXe-filled capillaries, of Teflon, Polyethylene or hydrogen-containing Teflon (Tefzel).

Neutron detection efficiencies for plain LXe, Teflon-capillaries and Tefzel-capillaries converters were about 20% over the entire energy range. In polyethylene capillaries converters the neutron detection efficiency was about 10% at 2 MeV and increased up to about 20% at 14 MeV. Detection efficiencies of gammas in Teflon, Tefzel and polyethylene converters were ~35%. The plain-LXe converter provided the highest gamma-ray detection efficiency, of ~40-50% for 2-15 MeV energy range.

Optimization of LXe-filled Tefzel capillary dimensions resulted in spatial resolution of ~1.5mm (FWHM) for neutrons and up to 3.5 mm (FWHM) for gamma-rays.

Simulations of radiographic images of various materials using two discrete energy gamma-rays (4.4 MeV and 15.1 MeV) and neutrons in broad energy range (2-10 MeV) were performed in order to evaluate the potential of elemental discrimination.

## 1. Introduction

Gamma and fast-neutron imaging technologies are being applied for investigating the content of aviation- and marine-cargo containers, trucks and nuclear waste containers (see for example Ref [1, 2]). High energy X-ray or Gamma radiographic inspection methods, such as Dual Energy Bremsstrahlung Gamma Radiography (DEBG) or Dual-Discrete-Energy Gamma Radiography (DDEG) [3], in which two discrete gamma rays at energy of 4.4MeV and 15.1MeV are used for special nuclear materials (SNM) detection, can provide high-resolution images of shape and density, and also some selectivity between high-Z elements. Fast-neutron imaging methods, such as Fast-Neutron Resonance Radiography (FNRR) [4], utilize a broad spectrum of fast neutrons in range of 2-10 MeV to provide a sensitive probe for identifying low-Z elements such as H, C, N and O, which are the main constituents of explosives and narcotics. In addition FNRR provides a mean for identifying the type of the explosive by determination of the density ratios of its main constituent elements [3].

The requirements from a radiographic gamma-ray and neutron detection system for contraband detection are: 1. spatial resolution of 3-5mm (FWHM), required for the detection of thin sheet explosives and small quantities of SNM; 2. detection efficiency of 20-30%; 3. capability of performing energy spectroscopy; 4. high counting rate capability of several MHz; 5. large area (~0.5m$^2$).

The general trend of future research projects in this area is to improve existing screening techniques by more efficient detection of contraband, including explosives, illicit drugs, illegal imports, weapons and special nuclear materials (SNM) [1, 5]. An inspection system featuring both FNRR and DDEG techniques will combine the capability of low-Z objects detection and substance-identification with high-Z selectivity. A Time-Resolved-Event-Counting-Optical-Radiation (TRECOR) detection system which combines imaging and energy spectroscopy of both fast neutrons and high-energy gamma-rays was developed in recent years by the German-Israeli Science and Technology (GIST) collaboration [6, 7, 8]. TRECOR utilizes plastic scintillator and LYSO crystals for the respective imaging of fast neutrons and gamma-rays. The neutron and gamma-ray spectroscopies are performed by the time-of-flight (TOF) technique and pulse-height analysis, respectively.

Recently, a new detector concept for combined imaging and spectroscopy of fast-neutrons and gamma rays, <u>in the same detection medium</u>, was proposed [9] and is shown schematically in figure 1. It encompasses a liquid-xenon (LXe) scintillator, in plain-volume or contained in "fiber-like", 1 mm in diameter capillaries with good UV-photon transport properties (e.g. made of Teflon). The scintillation-light ($\lambda\sim 178$ nm) created in the LXe converter, induced by neutron or gamma interactions with the Xe atoms (resulting in nucleon or electron recoils, respectively), propagates along the converter (e.g. within capillaries by total internal reflection); it is detected by a position-sensitive gaseous photomultiplier (GPM) [10] through a UV-transparent window. The photoelectrons, induced by UV-photon conversion on a CsI photocathode deposited on the top electrode of the GPM, are multiplied by successive gas electron multipliers, e.g. here cascaded THGEM electrodes [11-14]. The properties of cryogenic THGEM-GPMs and other GPM configurations with CsI-coated THGEM followed by a PIM (Parallel Ionization Multiplier) and a MICROMEGAS (MICROMEsh GAseous Structure) multipliers, coupled to LXe converters, were described elsewhere [15, 16]. The localization of the interacting radiation in the LXe converter is derived from the center of the gravity (CG) of all photons detected by the GPM.

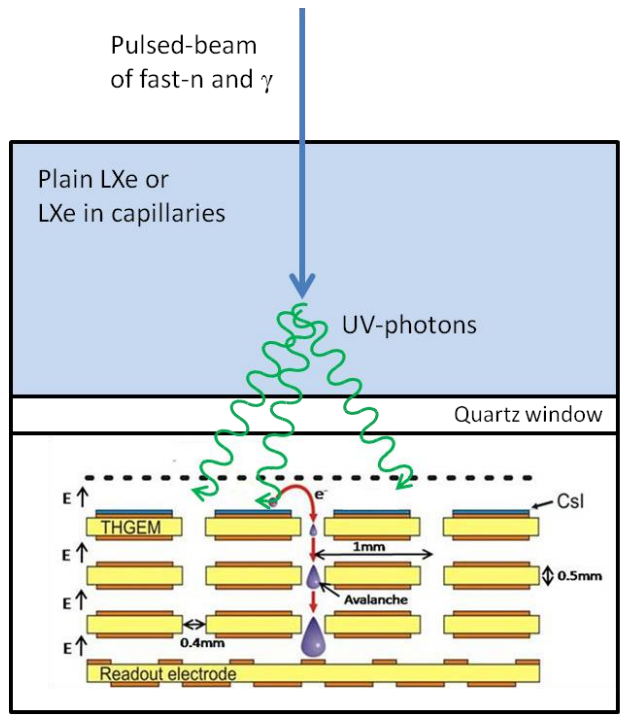

**Figure 1: Schematic drawing of the combined gamma & fast-neutron imaging detector concept. The interaction of radiation with LXe induces a fast scintillation-light flash. These UV photons are detected by a reflective CsI photocathode, coated on gas-avalanche electron multiplier – here a triple-THGEM, and followed by a segmented readout anode. The LXe sensitive volume and the GPM are separated by a UV-transparent quartz window.**

A previous simulation study [9], was made with 51 mm long, 1 mm inner diameter, closely packed LXe-containing Teflon capillaries. Teflon was chosen due to its low-reflective-index ($n_{teflon}$=1.34 vs $n_{Xe}$=1.61 @178nm), assuring efficient scintillation-light propagation along the capillaries by total reflection. For this converter configuration, detection efficiencies of around 20% and 30%, and spatial resolutions (CG) of about 10 mm and 4 mm (FWHM) for neutrons and gammas, respectively, were calculated in the relevant energy range of 2 to 14 MeV. For neutrons, the relatively poor spatial distributions of the CG of the detected light, were due to their elastic scattering resulting in scintillation-light occurring in several distant capillaries simultaneously; the relatively small number (~90) of detected photoelectrons per neutron event, also caused large uncertainties in the radiation localization. In the case of gammas, the resolution is dictated by the long-range of gamma-ray induced electrons/positrons (mainly Compton and pair production processes).

The present simulation study, using GEANT4 toolkit (version 9.3.2) [17], is aimed at finding converter configurations with improved localization resolution and detection efficiency. We investigated these parameters in four radiation-converter configurations consisting of: plain LXe medium and LXe-filled capillaries; the latter made of Teflon ($C_2F_4$), Tefzel ($C_4F_4H_4$) [18] or Polyethylene (($C_2H_4)_nH_2$). Hydrogen-rich capillaries, such as polyethylene or Tefzel, allow efficient transfer of neutron energy close to the point of interaction, thus improving the spatial resolution. Polyethylene is a hydrogen-rich material but it poorly reflects UV photons. In order to collect the photons efficiently by total internal reflection, the inner surface of polyethylene capillaries must be coated with Teflon (e.g.10 μm in our simulations). Tefzel is also hydrogen-rich and additionally reflects effectively UV photons in the considered wavelength [19];

## 2. Neutron and gamma-ray interaction with LXe, and capillary materials

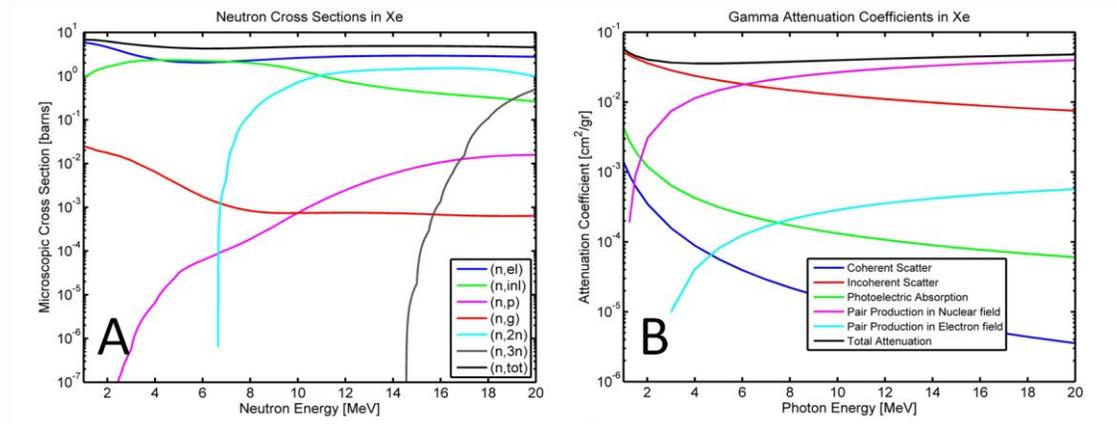

**Figure 2:** Neutron's cross sections (fig. A) and gamma's mass attenuation coefficients (fig B) of natural Xe for the energy range 1-20 MeV. In fig A (n,el) stands for neutron elastic scattering, (n,il) stands for neutron inelastic scattering, (n,p) stands for neutron-proton reaction, (n,g) stands for neutron capture reaction, (n,2n) and (n,3n) stand for neutron-2 neutrons and neutron-3 neutrons reaction and (n,tot) stands for the total cross section. Neutron's data is taken from [20], gamma's data is taken from [21].

2.1 **Neutron and gamma-ray interaction in LXe**

The neutron cross-sections [20] and gamma-ray mass attenuation coefficients of Xe [21] are shown in figure 2. The prominent neutron reactions with Xe, in the relevant energy range of 2-20 MeV, are elastic and inelastic scattering. Much less probable but still important for energy deposition in the converter are the neutron capture and (n,p) reactions.

Neutrons that undergo inelastic collisions with xenon, deposit part of their energy to excitation of the nucleus, producing γ-rays as it de-excites. The energy deposited by these gamma-rays will add to the total energy deposited by the neutron in LXe.

The dominant processes of gamma-ray interaction in LXe above 2 MeV are Compton scattering (incoherent scattering in figure 2) and pair production. Both processes produce energetic electrons or electrons/positrons. Above approximately 7 MeV the pair production process becomes dominant. All these processes were accounted for in our simulations.

## 2.2 Neutron and gamma-ray interaction in capillary materials

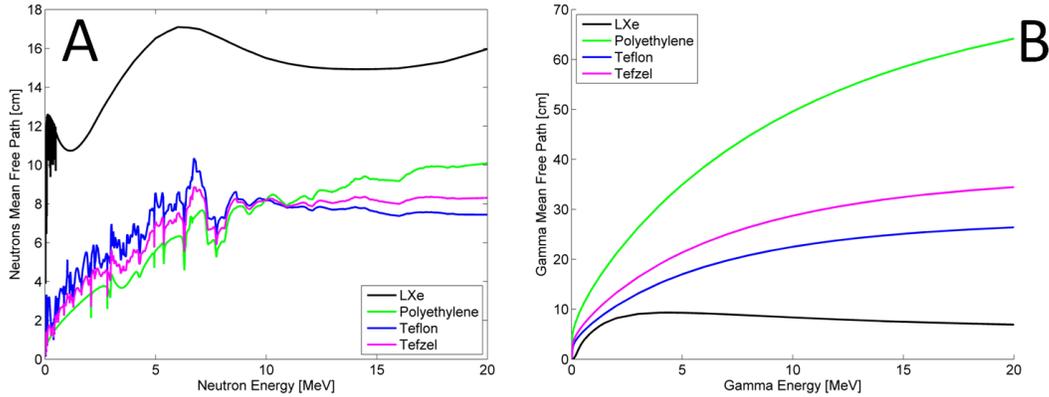

**Figure 3: Mean free path of neutrons (figure A) and gammas (figure B) in the various radiation-converter materials (LXe, Polyethylene ($C_2H_4$), Teflon ($C_2F_4$) and Tefzel ($C_4F_4H_4$)). The neutron and gamma mean-free-path values were calculated from data taken from [20] and [21], respectively.**

Figure 3 shows the mean-free-path values for fast neutrons and gamma-rays in capillary materials such as Polyethylene, Teflon and Tefzel. Xe data is also shown for comparison. These values, for neutron and gamma, were calculated from data taken from [20] and [21], respectively. Although the dimensions of the LXe converter are smaller than the neutron mean free path in LXe, there is a small probability for multiple neutron scattering within the converter volume. The neutron's mean free path in all of the shown light materials, is shorter than in LXe. Hence, including these materials within the radiation-converters (e.g. in the form of capillaries) will increase the probability of interaction and lead to an efficient energy deposition of neutrons close to their impinging point. In this manner, the neutron's spatial resolution may improve.

For gamma rays, on the other hand, the mean free path in LXe is much shorter than the mean free path in all the light materials considered. Hence, the gamma's mean free path in the converter will be influenced mainly by the LXe and no improvement of the gamma's spatial resolution is expected due to presence of light materials.

## 3. The simulated detector configurations

The following detector configurations were investigated through simulations:

a) **Plain-LXe converter:** A 51 mm thick LXe sensitive volume, viewed via a 10mm thick quartz UV-window by the THGEM-GPM. The CsI photocathode is deposited on the THGEM's top face (reflective mode), located 5 mm away from the UV-window; the GPM is filled with Ne:10%CF$_4$ at 1 atm. A GEANT4 snapshot of this concept is depicted in figure 4 (right).

b) **Teflon capillaries, filled with LXe (studied at [9]):** Array of 50 mm long and 1 mm diameter holes, drilled at an hexagonal pattern with a pitch of 1.2 mm in a block of Teflon and filled with LXe. Teflon was chosen due to its low refractive index compared to that of LXe, at 178 nm ($n_{Teflon}$=1.34 vs $n_{LXe}$=1.61). Total internal reflection guides part of the radiation-induced scintillation photons towards the 10 mm UV window (UV-grade synthetic quartz; $n_{quartz}$ =1.6). A 1mm gap is maintained between the capillaries and the window to allow for LXe circulation. The GPM layout is the same as in (a). A GEANT4 snapshot of this concept is depicted in figure 4 (left).

c) **Tefzel capillaries, filled with LXe:** Array of 50 mm long and 1 mm diameter holes arranged in the same geometry as in (b), drilled in Tefzel (index of refraction $n_{Tefzel}$=1.5 at178 nm [19]), filled with LXe.

d) **Polyethylene capillaries, coated with thin Teflon layer and filled with LXe:** Array of 50 mm long and 1 mm diameter holes arranged in the same geometry as in (b), drilled in Polyethylene. The inner walls of the Polyethylene capillaries are coated by 10 μm of Teflon, to gain total internal reflection (not possible with polyethylene). The high Hydrogen content is supposed to reduce neutron multiple scattering to distant capillaries.

We assumed photon transmission through the capillary holes by a specular total internal reflection [22]. The incident neutron and gamma-ray beams had an area of 1.4 x 1.4 mm$^2$, impinging at the center of detector's active volume. Our calculations included all steps in the chain of signal development: the total deposited-energy distribution in LXe (by electron or Xe nuclei recoil, respectively, for gamma and neutron interactions); the total scintillation photon-yield and its spatial distribution inside the LXe volume; the amount of UV-photons transported along the capillaries and through the UV-window; the photon collection efficiency on the photocathode surface; the spatial distribution of the photoelectrons emitted from the photocathode and its center of gravity (defining the spatial resolution of the detector system) and the detection efficiency.

The simulation results for setup (b) have been published elsewhere [9]: in this work they serve as a means for comparison and for further converter optimization.

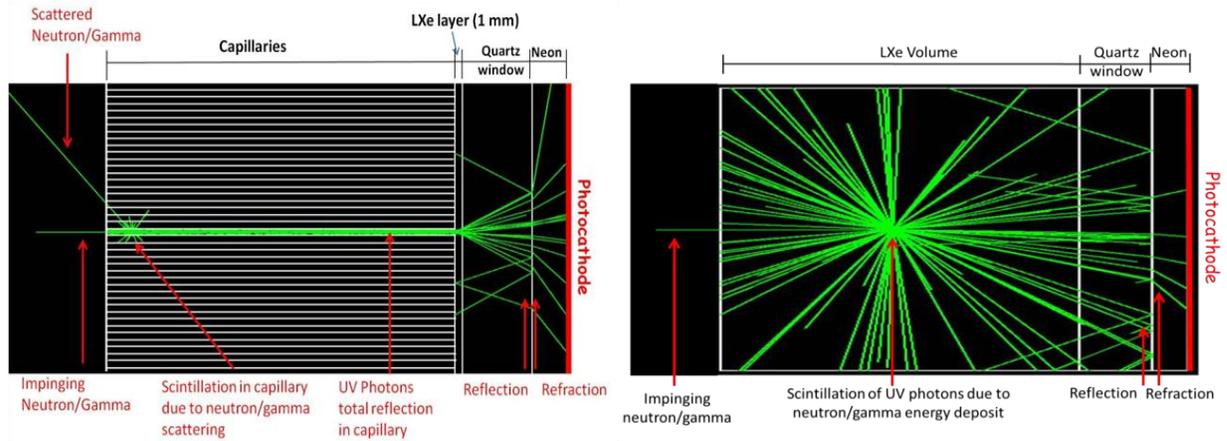

**Figure 4: Snapshots of GEANT4 simulation runs (side view): the green lines are tracks of neutral particles (either incident neutron/gamma or UV-photons). Left figure: LXe-filled capillaries (50 mm long and inner diameter of 1 mm) made of Teflon, Tefzel or Polyethylene with Teflon coating; a LXe-filled 1mm gap; a 10mm thick UV-window; a 5mm Ne-based gas gap and a photocathode surface (of the GPM). (Taken from [9]). Right figure: plane 51 mm long LXe volume; a 10mm thick UV-window; a 5mm Ne-based gas gap and a photocathode surface.**

## 4. Simulation results

### 4.1 Deposited Energy Spectra

Figure 5 and figure 6 depict the spectra of deposited energy in the various LXe radiation-convertor setups, computed for gamma and neutron interactions in the relevant energy range 2-14 MeV.

4.1.1 Gamma-ray spectra

For gamma-ray interactions in plain LXe volume (figure 5-A) one can clearly see an intense full energy photo-peak, as well as single- and double- escape photo-peaks from the dominant pair-production process. A small, but visible, peak at 511 keV on the low energy side of the gamma spectra originates from pair production and annihilation processes in the 10 mm thick quartz window. In comparison to a plain LXe volume, the introduction of capillaries (of the geometry described above) causes the following effects: Only 55% of the incident beam particles interact directly with LXe. Gamma-ray induced electrons/positrons created in LXe will deposit only part of their energy within the active volume. Incident particles that interact directly with capillary material create predominantly Compton electrons, of which some reach the active LXe volume and deposit there part of their energy. Hence, we do not see the full or escape photo-peaks except in the case of low-energy gamma-rays where the probability to stop an electron/positron within the active LXe is still relatively high.

4.1.2 Neutron spectra

Figure 6 -A depicts energy spectra from neutron interaction in plain LXe. In a single elastic collision with Xe nucleus, the neutron can transfer not more than about 3% of its energy to Xe. Thus, for a relatively small volume of LXe, most of the elastic scattering events will deposit small amount of neutron energy per interaction. For example for 2 MeV neutrons the maximum energy transferred to Xe nucleus is about 60 keV. However, one can observe that the neutron spectrum extends up to its maximum energy. This is due to inelastic neutron collisions, where the resulting gamma-rays can add their energy to Xe recoil energy. In rare cases, the neutron may deposit higher energy than its incident energy. This may happen when inelastic collisions are followed by a neutron capture reaction. In such cases the absorption of the capture gamma-rays will add to the energy of the neutron.

The introduction of Teflon capillaries did not change significantly the absorbed energy spectrum. However, for capillaries made of materials that contain hydrogen, such as Tefzel or polyethylene, the neutron spectra extended to higher values due to the contribution of knock-on protons, which may receive large fraction of neutron energy in a single collision.

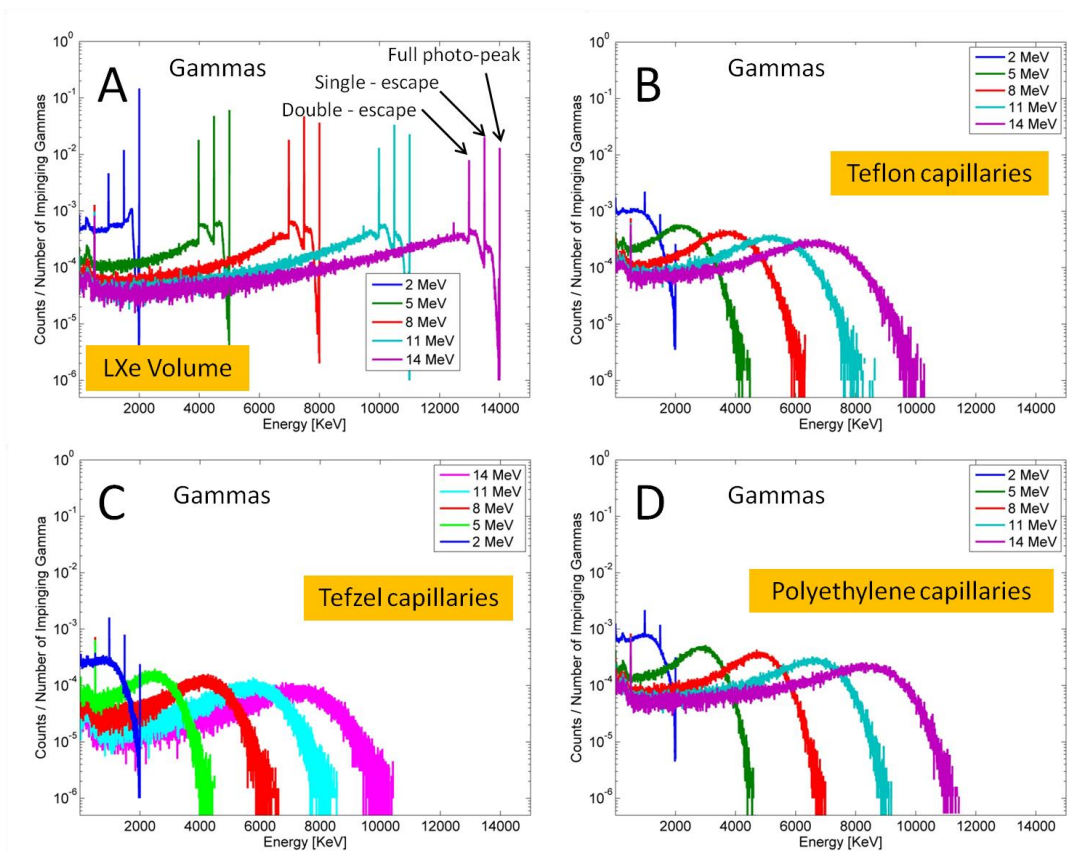

**Figure 5:** Computer-simulated spectra of deposited energy in LXe by gamma photons for a) plain-LXe converter, b) Teflon capillaries, c) Tefzel capillaries and d) Polyethylene capillaries. The simulations were made for impinging-radiation energies of 2, 5, 8, 11 and 14 MeV.

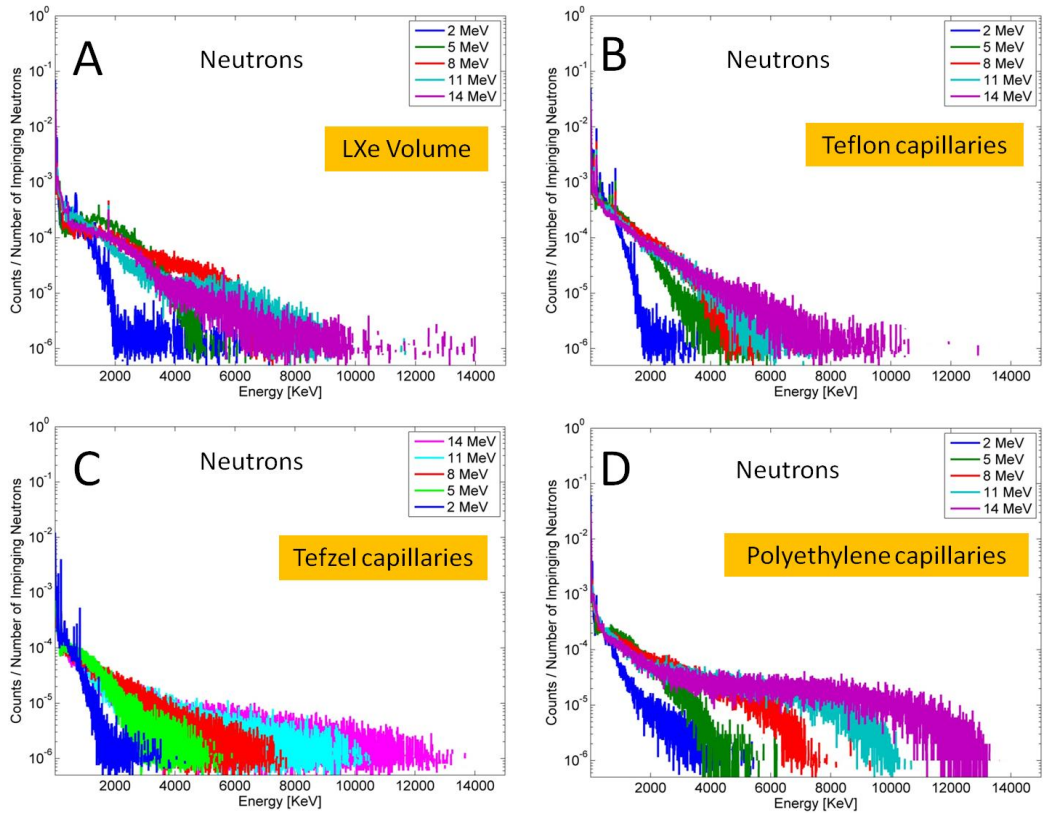

**Figure 6: Computer-simulated spectra of deposited energy in LXe by neutrons for a) plain-LXe converter, b) Teflon capillaries, c) Tefzel capillaries and d) Polyethylene capillaries. The simulations were made for impinging-radiation energies of 2, 5, 8, 11 and 14 MeV.**

The average gamma-ray and neutron energies deposited in the active volume and available for light production are shown in table 1. It includes all possible modes by which the incident particle can leave its energy to scintillation in LXe.

**Table 1: Average deposited energy, in the various detector setups, for selected gamma and neutron beam energies.**

| Energy of impinging beam [MeV] | Average deposited energy in LXe volume setup [keV] | | Average deposited energy in Teflon capillaries setup [keV] | | Average deposited energy in Tefzel capillaries setup [keV] | | Average deposited energy in Polyethylene capillaries setup [keV] | |
|---|---|---|---|---|---|---|---|---|
| | gamma | neutron | gamma | neutron | gamma | neutron | gamma | neutron |
| 2 | 1310 | 220 | 700 | 240 | 740 | 200 | 820 | 170 |
| 5 | 3590 | 900 | 1820 | 620 | 1970 | 620 | 2230 | 620 |
| 8 | 5930 | 1100 | 3020 | 760 | 3260 | 840 | 3680 | 1020 |
| 11 | 8100 | 750 | 4150 | 720 | 4510 | 910 | 5020 | 1340 |
| 14 | 10030 | 730 | 5170 | 810 | 5640 | 1120 | 6200 | 1740 |

## 4.2 Detection Efficiency

The detection efficiency is defined here as the number of particles (neutron/gamma) interacting in the LXe sensitive volume, resulting in at least one photoelectron, detected by the GPM, normalized to the total number of particles impinging on the detector. The scintillation light yields of 8.8 photons/keV and 20 photons/keV, for neutron-induced nuclear recoils and gamma-induced electron recoils, respectively, used in our simulations were taken from [23]. The detection efficiency of gammas and fast neutrons, computed for the different converter variants over the relevant energy range, are shown in figure 7. Neutron detection efficiency of the early version of TRECOR, the TRION detector (30 mm thick scintillation-screen + intensified CCD) [24, 25], is also shown for comparison. The detection efficiency is rather constant over the whole energy range, of the order of 20% and 30%-40% for fast neutrons and gammas, respectively. For gammas, due to the large energy deposition by Compton electrons and their resulting high scintillation yield (see figure 5 and table 1), the detection efficiency is equal to the conversion efficiency; namely, every gamma interacting in LXe generates at least one photoelectron detected by the GPM.

Detection efficiencies of gammas in Teflon and Tefzel converters are roughly the same. The low density of polyethylene causes some reduction in gamma-ray efficiency. The plain-LXe converter provided the highest detection efficiency, e.g. of ~45-55% for 2-14MeV gammas.

For neutrons, a significant number of events deposit very small amount of energy in LXe (see figure 6 and table 1). Furthermore, the small number of scintillation photons emitted per keV of deposited energy by neutron-induced Xe recoils will result in a lower detection probability. Detection efficiencies of neutrons in the Teflon-capillaries and the plain LXe volume configurations are roughly the same, over the whole energy range.

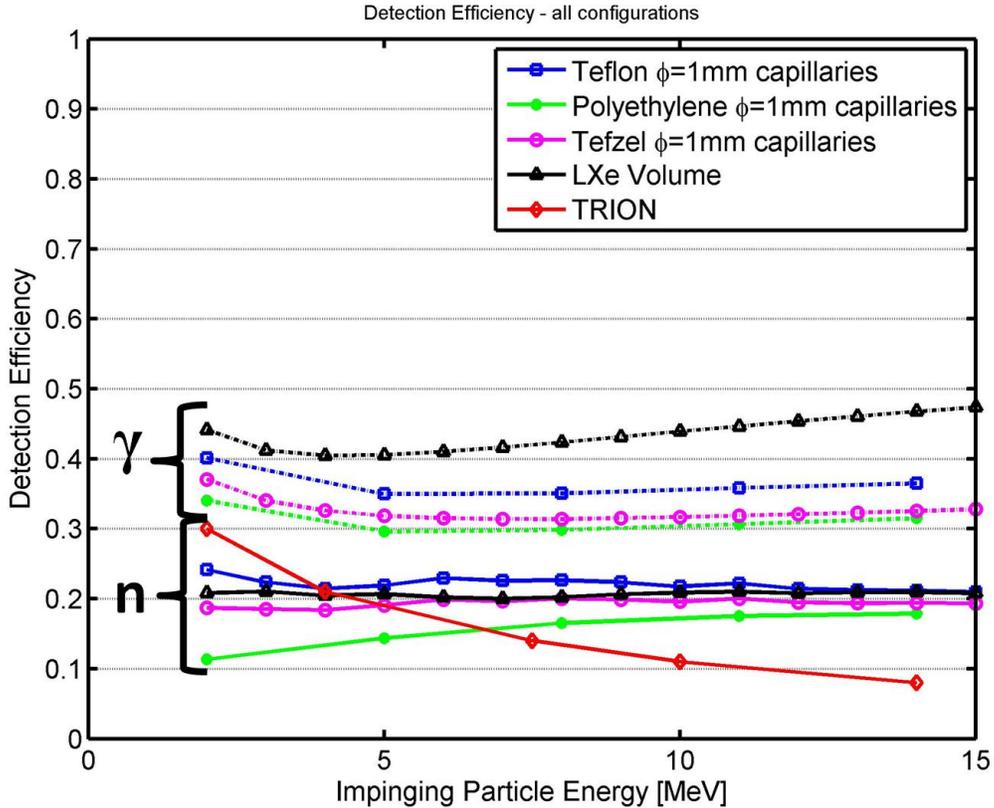

**Figure 7: Detection efficiency as a function of energy of the impinging neutrons (solid lines) and gammas (dashed lines) for the detector of figure 1 with a plain-LXe convertor and that of Teflon, Tefzel and Polyethylene capillaries. Converter thickness: 50mm. Neutron detection efficiency, as calculated in [24] for TRION detector, is shown for comparison.**

One can notice that the neutron efficiency for polyethylene capillaries is significantly lower for low neutron energies. It is thus interesting to investigate the contribution of the knock-on protons to the detection efficiency. The proton-induced scintillation yield is about 8-fold higher (per keV of deposited energy) compared to that of Xe nuclear recoils [23] (in addition to a higher maximum energy transfer, since the proton deposits all its energy within the medium (see figure 8)); hence, the scintillation light collection efficiency and the resulting detection efficiency are expected to improve. For that purpose, in the analysis of the simulation results, we separated the deposited energy induced by proton originating from the Polyethylene capillary walls (and passing through 10 μm of Teflon coating), from that of the recoil Xe nuclei (see figure 8).

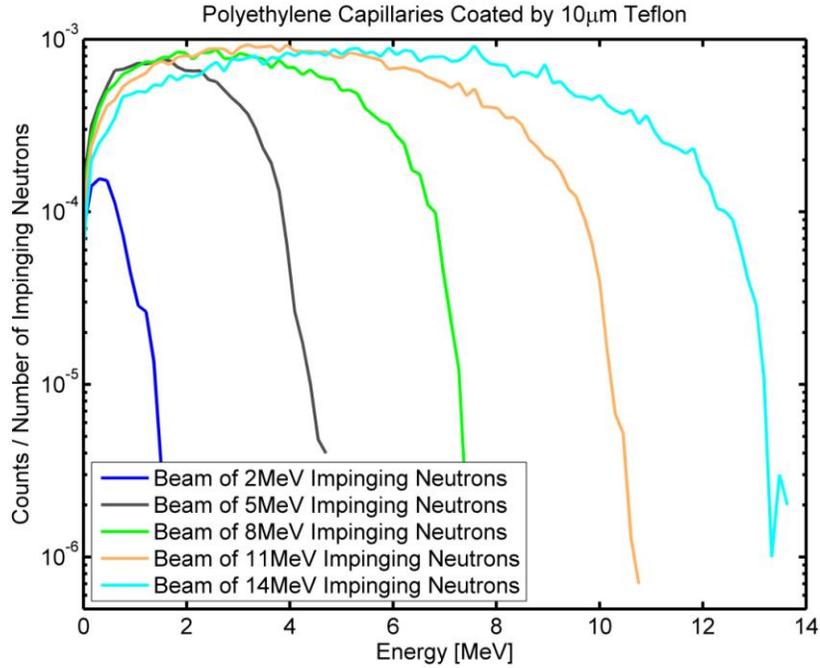

**Figure 8: Energy distributions of protons, released from the Polyethylene wall by interacting neutrons (2, 5, 8, 11 and 14 MeV), penetrating into the LXe capillaries.**

Table 2 summarizes the efficiencies (energy integral of the distributions figure 8) due to protons only. As can be observed, most of the knock-on protons are trapped in the capillary wall; therefore, their contribution to the total detection efficiency is insignificant. Our conclusion is that the reduction of efficiency at low neutron energy is caused by the lower average deposited energy available for production of light (see table 1)

**Table 2: Number of protons penetrating into LXe normalized to the number of impinging neutrons.**

| Energy of impinging neutrons [MeV] | Number of detected protons/ Number of impinging neutrons [%] |
|---|---|
| 2 | 0.08 |
| 5 | 1.37 |
| 8 | 2.66 |
| 11 | 3.98 |
| 14 | 4.94 |

## 4.3 Spatial resolution

The reconstruction of the original position of the impinging particle is obtained by calculating the center of gravity (CG) of the cloud of photoelectrons detected by the position-sensitive GPM. Examples of single-event snapshots, simulated in GEANT4 runs, are illustrated in figure 9 from the detector's front-side (as seen by the photocathode); they show the results of single neutrons and gammas impinging on the LXe sensitive volume, yielding nuclear and electron recoils, respectively, recorded through their resulting scintillation light. The examples of the radiation-induced recoils are shown in a plain LXe-volume and in capillaries drilled in Teflon and Polyethylene, filled with LXe (represented as 1 mm diameter circles). The initial position of the impinging particle is indicated by white arrows. The green lines and spots are neutral particles (i.e. neutrons, gammas or the emitted UV-photons). The transport of the UV-photons, in the plain LXe volume, was suppressed, in order to enable better viewing.

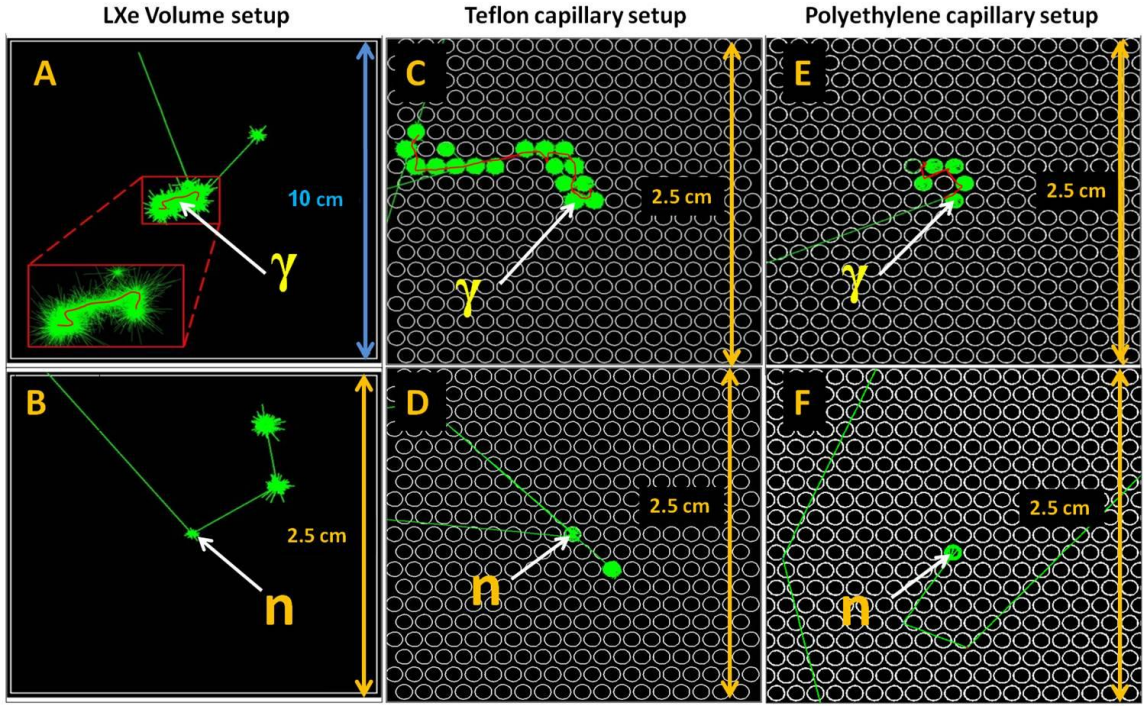

**Figure 9: Examples of snapshots of typical GEANT4 runs, seen from the detector's front-side, for impinging neutrons and gamma-rays in a plain LXe converter (figure A, B), LXe in Teflon capillaries (figure C, D) and LXe in Polyethylene capillaries (figure E, F). The scintillation light (green spots) is created by radiation-induced recoils stopped in LXe. The Compton-electron tracks appear in red. The white arrows indicate the impact locations. While the UV-photon transport in confined within capillaries, their transport in the plain LXe volume was suppressed to enable better viewing. See text.**

As illustrated in figure 9, the uncertainty of reconstructing the original position of the impinging gamma radiation is mostly due to the long range of Compton electrons. The maximum energy of the ejected Compton electron varies from 88% to 98% of the initial gamma energy for gamma ray energy range of 2-15 MeV. Their range in LXe reaches several mm; they induce large scintillation-photon yields in several contiguous capillaries or within a few mm of LXe volume. This is well illustrated in upper figures of figure 9; the interactions of the Compton electron (in red) and the scintillation photons (in green) occur within the capillaries and in the LXe volume along the Compton-electron track.

Figure 9-B and figure 9-D illustrate cases in which the neutron is scattered a few times at distant points (three points in fig B and two capillaries in fig D). Reconstruction of the neutron-induced interaction position, for these multiple-scattering processes in the LXe, are difficult due to the large spread of the scintillation light. On the contrary, figure 9-F shows an ideal situation where the impinging neutron is directly scattered within the capillary, then scattered again few times in the polyethylene support, losing its energy without scintillation. Reconstruction of the original position of the impinging neutron in this case is straightforward.

Figure 10 illustrates the spatial distribution of the detected photoelectrons on the GPM, for three of the radiation-converter setups. The color scale relates to the total number of detected photoelectrons. The figure depicts the photoelectrons cloud resulting from single events, its calculated center-of-gravity (cg) and the original impact location (i) (as explained below).

In the case of neutrons, the geometrical spread of the detected photoelectron distributions in the detector with LXe in Teflon capillaries and in the plain LXe volume converter is larger than that with the LXe-filled Polyethylene capillaries one. This is due to a larger number of multiple-scattering interactions in the former configurations.

In the case of gamma irradiation, the geometrical spread of the detected photoelectron distributions in the plain LXe converter is the largest one; nevertheless, the original position of the interacting particle is calculated with a rather small uncertainty due to the large number of photoelectrons.

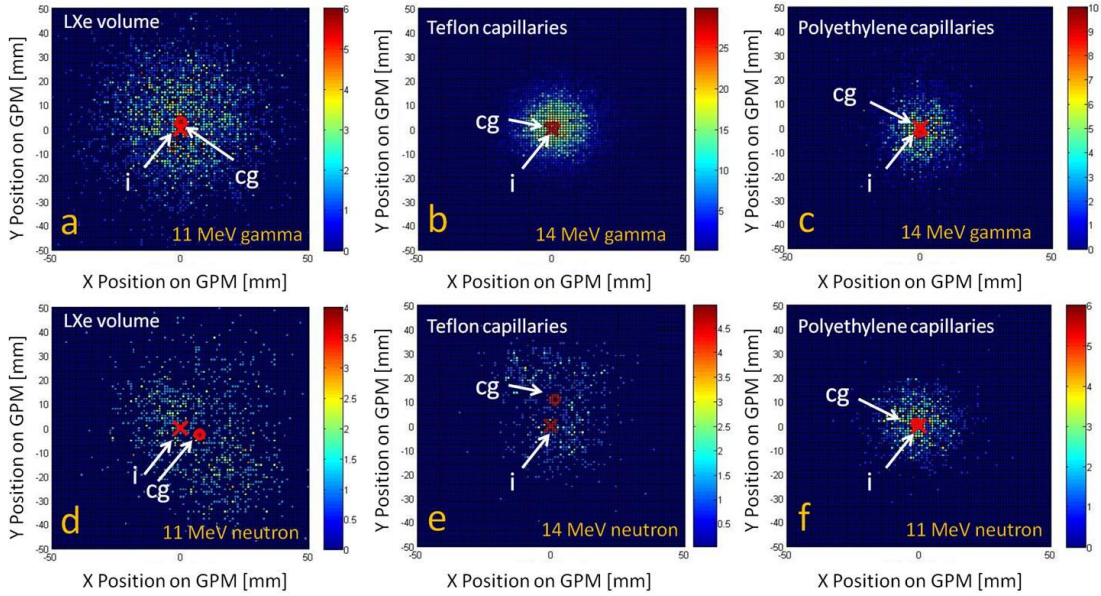

**Figure 10:** Examples of 2D spatial distributions of the photoelectrons on the GPM's photocathode (detected photoelectrons), simulated for three LXe converter geometries (indicated in the figure), for gamma-rays (a, b and c) and neutrons (d, e and f). The color indicates the number of detected photoelectrons in each position; (i) is the original radiation-impact location; (cg) is the computed center-of-gravity of the detected photoelectrons cloud.

The detector's spatial resolution was obtained in each configuration by computing event-by-event, the centers-of-gravity (cg) of the detected photoelectrons distributions on the GPM's photocathode surface. The distributions, resulting from $2\text{-}20\cdot 10^6$ impinging particles, are shown in figure 11. Summary of the FWHM values of the cg distributions is shown in figure 12.

For the neutrons, the simulated spatial distributions, and consequently the resolutions, depend on the capillaries' substrate material; this is due to the large differences between the neutron total cross sections in Xe, Teflon, Tefzel and Polyethylene, affecting the neutron's mean-free-path in the converter and the amount of energy transferred to scintillation in LXe. On the other hand, the gamma-induced spatial distribution is almost independent on the capillaries' material, since gamma rays interact mostly with LXe.

Therefore, one may conclude that, while no advantages are reached with the use of capillaries as scintillation-light guides in term of spatial resolution for gammas, they largely improve the spatial resolution for neutrons by more efficient transfer of neutron energy near the point of neutron interaction.

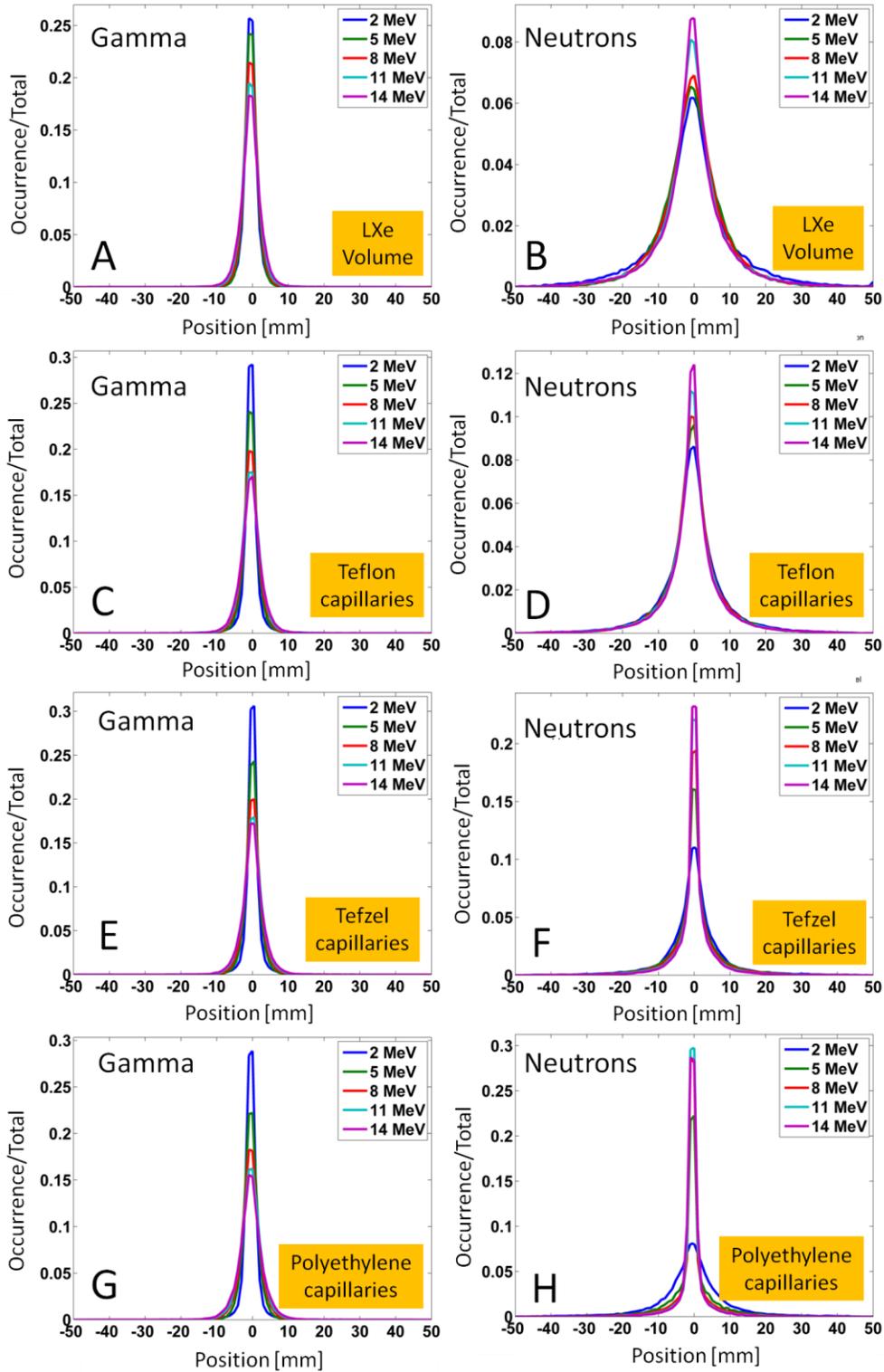

**Figure 11:** Computed centre-of-gravity distributions of photoelectrons from the GPM's photocathode resulting from gamma- and neutron-induced scintillation light in detectors with different radiation converter configurations. Converter thickness: 50mm; energy range: 2-14 MeV. The distribution areas are normalized to 1.

As expected, for neutrons, the narrowest spatial distribution was obtained with the Polyethylene or Tefzel capillaries (FWHM of ~2.5mm), due to higher energy transfer to the capillary materials close to the point of interaction and higher photons statistics in case of scintillation from knock-off-protons. The broadest spatial resolution was obtained in the plain LXe Volume (FWHM of ~8mm).

The large number of photoelectrons emitted in most gamma interactions, and the shorter Compton–electron range in LXe compared to that of the neutron's range, resulted in narrower spatial distributions for gamma. Furthermore, the small cross sections for gamma interaction in Teflon, Tefzel and Polyethylene, make these materials "transparent" to gammas, hence similar spatial resolutions were obtained for gamma in all four different detector configurations (FWHM of 3-5mm).

The deterioration of the spatial resolution with the gamma energy is caused by the increase of the Compton-electron range [26].

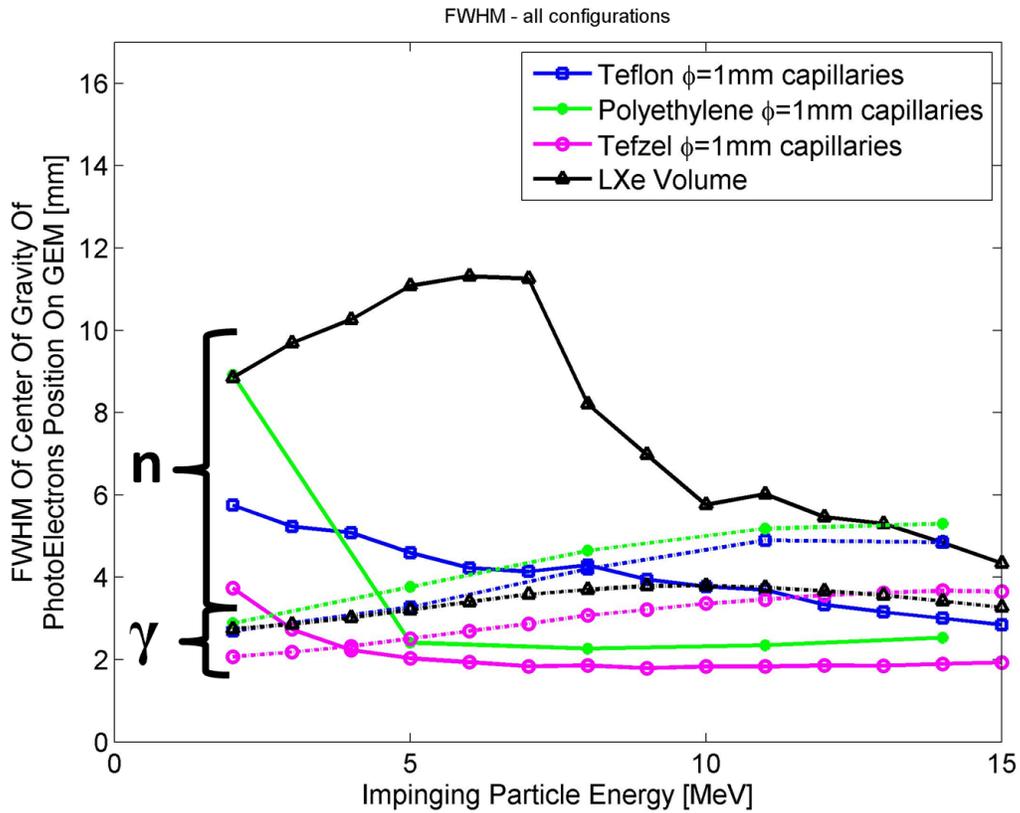

**Figure 12: Computed FWHM values of the photoelectrons' center-of-gravity distributions on the GPM's photocathode (of figure 11), as a function of energy of the impinging neutrons (solid lines) and gammas (dashed lines) for detectors with different, 51mm long radiation converter configurations (indicated in the figure).**

## 4.4 Capillary - dimensions optimization

In paragraph 4.2 (figure 7) we demonstrated that detection efficiencies of neutrons in the Tefzel- or Polyethylene capillaries and in the plain LXe volume setups are roughly the same, over the whole spectrum. On the other hand, the spatial resolution, obtained for fast-neutrons in the Tefzel- or Polyethylene capillaries setup is considerably narrower compared to that of the plain LXe volume setup - due to efficient neutron energy losses by collisions with H atoms.

The relatively low melting temperature of Polyethylene probably prevents coating the inner walls of capillaries by a Teflon reflector. Hence, the best practical detector configuration, based on Geant4 simulations, is one with Tefzel capillaries. The Tefzel capillaries' dimensions (e.g. inner and outer radii) were optimized for neutron and gamma irradiation, by Geant4 simulations, performed on commercially available capillary dimensions [27]. The results of the FWHM values of the CG distributions of neutron- and gamma-induced photoelectrons' position on the UV detector, and its detection efficiency, are shown in figure 13 and table 3, respectively, for different capillary sizes and neutron and gamma energies. The optimal capillaries' dimensions (out of the existing selection in table 3) are an inner radius of 0.508 mm and outer radius of 0.794mm. For these dimensions, the calculated neutrons' detection efficiency is ~0.18 (2-10MeV neutrons) and that for gammas is ~0.3 (4.4 and 15.1MeV gamma). The calculated FWHM spatial resolution for fast-neutrons is ~1.5mm (for 2-10MeV neutrons) and that for gammas ~2.4mm (@4.4MeV) and ~3.5mm (@15.1MeV).

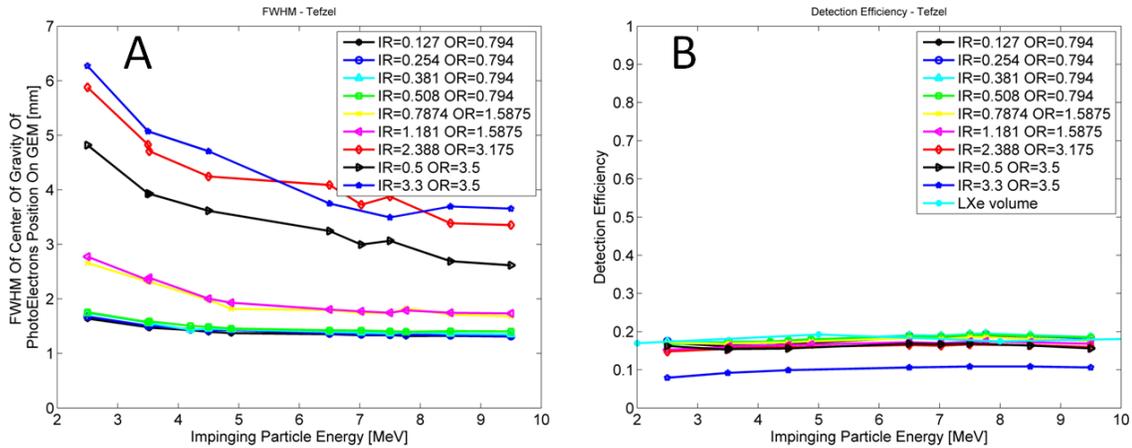

**Figure 13: FWHM of CG distribution of photoelectrons position on the GPM (figure A) and detection efficiency (figure B) for different neutron energies, calculated in the different capillaries sizes. In figure B, the detection efficiency of plain LXe volume, with no capillary, is shown for comparison.**

**Table 3: FWHM of CG distribution of photoelectrons position on GEM, and detection efficiency for the two calculated gamma energies, for different capillaries sizes. The best commercialy-available Tefzel's dimensions, in terms of detection efficiency, are in the gray line.**

| Inner Radius [mm] | Outer Radius [mm] | FWHM [mm] | | Detection Efficiency | |
|---|---|---|---|---|---|
| | | 4.4 MeV | 15.1 MeV | 4.4 MeV | 15.1 MeV |
| 0.127 | 0.794 | 2.7 | 2.9 | 0.24 | 0.19 |
| 0.254 | 0.794 | 2.6 | 3.1 | 0.25 | 0.21 |
| 0.381 | 0.794 | 2.6 | 3.2 | 0.27 | 0.25 |
| **0.508** | **0.794** | **2.4** | **3.5** | **0.29** | **0.30** |
| 0.7874 | 1.5875 | 2.7 | 3.6 | 0.26 | 0.25 |
| 1.181 | 1.5875 | 2.9 | 4.3 | 0.29 | 0.33 |
| 2.3876 | 3.175 | 3.6 | 5.1 | 0.26 | 0.33 |
| 0.5 | 3.5 | 2.6 | 3.1 | 0.20 | 0.19 |

### 4.5 Simulation of radiographic images – elemental differentiation

GEANT4 simulations were carried out in order to evaluate the expected performance, of both, the plain LXe volume radiation-converter and of Tefzel capillaries radiation-converter of the optimal dimensions (see section 4.4), for gamma and neutron radiography and material differentiation. The plain-LXe configuration was chosen as worst case scenario since its spatial resolution, particularly for neutrons, is the lowest. The Tefzel capillaries configuration was chosen as best case scenario since its spatial resolution, particularly for neutrons, is the highest. Nine objects (20x20x20mm$^3$ for gammas and thicker ones, 20x20x60mm$^3$ for neutrons) of various materials were considered: Lead, Tungsten, Uranium, Polyethylene, Graphite, Aluminium, PETN (an explosive), Iron and silicon; they were "irradiated" by a uniform discrete energy gamma beams obtainable from the $^{11}$B(d,n;γ)$^{12}$C reaction (4.4 MeV and 15.1 MeV, ~1600 gammas/mm$^2$ for each energy) and by neutrons (continuous spectrum of 2-10MeV, ~2800 neutrons/(MeV·mm$^2$). The transmitted radiation was "measured" by the GPM detector. The simulated gamma-ray pulse height spectrum is shown in figure 14 for a plain LXe volume and Tefzel converters. One can observed that compared to the absorbed-energy gamma-ray spectra (see figure 5-A and C) the peaks here are smeared due to a poor light collection. Nevertheless, the two gamma-rays are well separated and the pulse height spectrum can be used for material differentiation. Neutron spectra will be measured by TOF spectroscopy, which is expected to provide very good energy resolution.

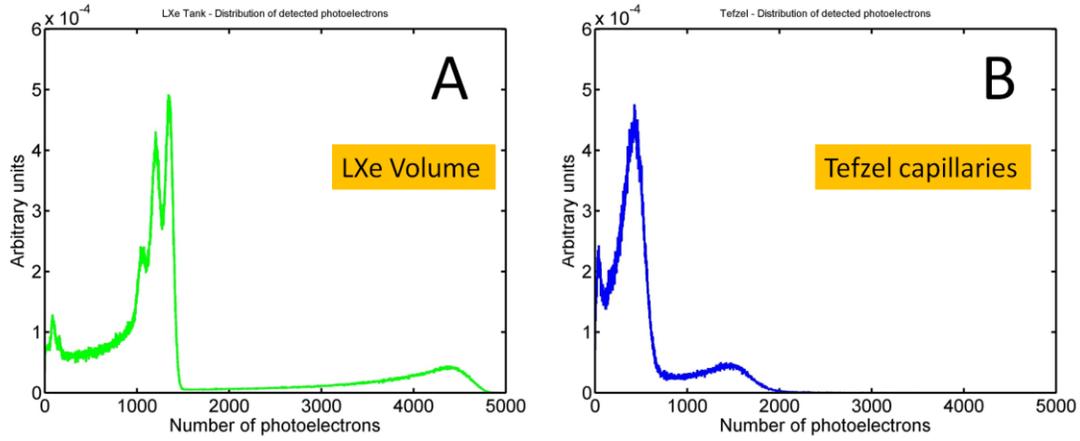

**Figure 14: Simulated gamma-ray (4.4 MeV and 15.1 MeV) pulse height spectrum for A-plain LXe volume and B-Tefzel converters.**

Typical simulated radiography results, with a detector with plain LXe converter and a detector with Tefzel capillaries convertor of the optimal geometry, are shown in figure 15, for selected gamma energy (4.4MeV), and selected neutron energy range (9-10MeV). The images were enhanced using Lucy-Richardson deconvolution and median filter algorithms, by MATLAB (version R2011b [28]).

The squared objects' shapes can be easily seen in the gamma radiography images with both convertors (figure 15-A and B), and in the neutron radiography with the Tefzel convertor (figure 15-D), while spread circle-like shapes were obtained in the neutron radiography with the plain LXe convertor (figure 15-C), due to the lower spatial resolution for neutrons in this convertor (see figure 12).

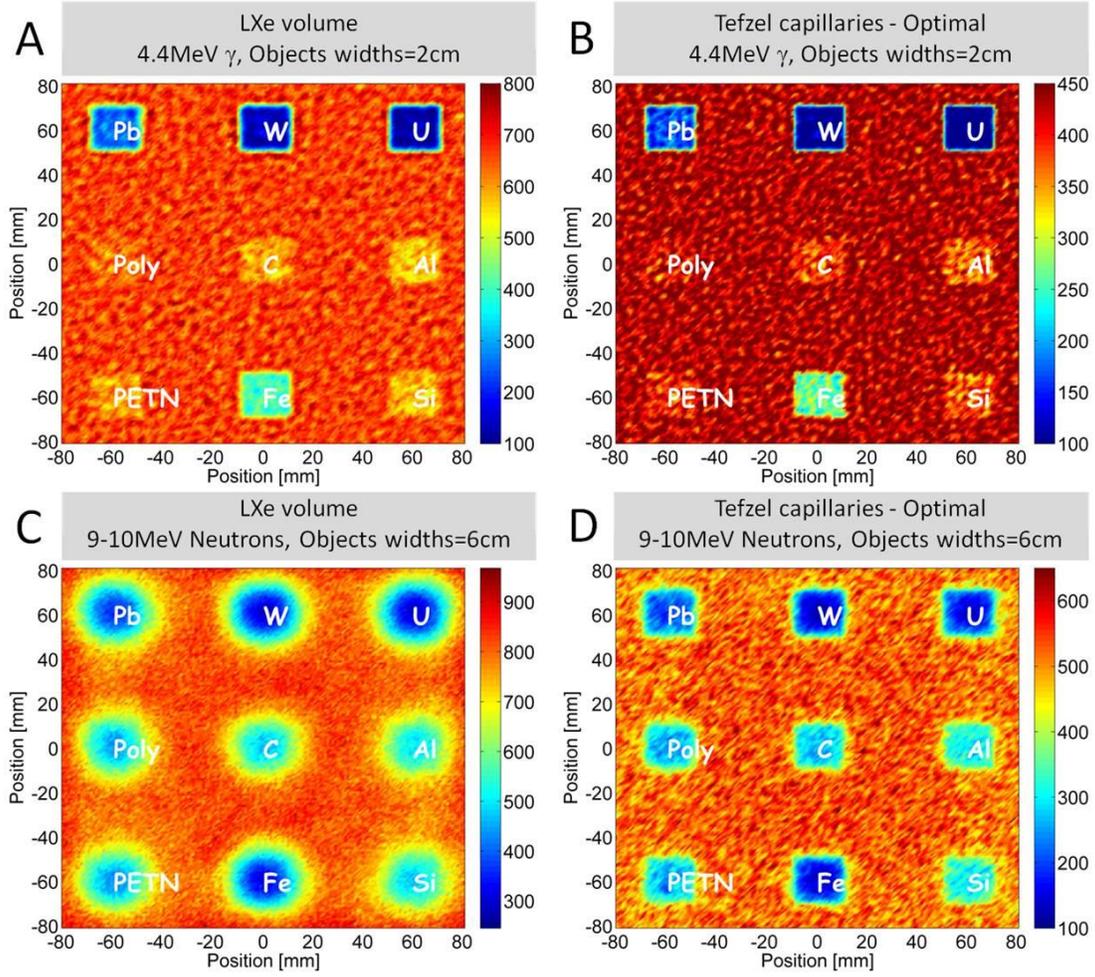

**Figure 15:** Typical simulated radiography results with a detector with plain LXe converter (figures A and C) and detector with Tefzel capillaries convertor of the optimal geometry (figures B and D), for objects of various materials. The figures show images for selected gamma energy (**4.4MeV**, figures A and B), and selected neutron energy range (**9-10MeV**, figure C and D). See text for details.

Simulations of material differentiation by dual-discrete-energy **gamma** radiography (DDEG [3]) were performed using $R_{value}$- defined, for each element, as the ratio between mass attenuation coefficients of gammas in the two selected energies (equation 1):

**(Equation 1)** $$R_{Value}(E_1, E_2) = \frac{\ln(I_\gamma(E_1)/I_\gamma^0(E_1))}{\ln(I_\gamma(E_2)/I_\gamma^0(E_2))} = \frac{\mu_\gamma(E_1)}{\mu_\gamma(E_2)}$$

Where here $E_1$ is 15.1MeV, $E_2$ is 4.4MeV, $I_\gamma^0$ is the impinging gamma flux, $I_\gamma$ is the transmitted flux and $\mu_\gamma$ is the mass attenuation coefficient. Materials with low, medium or high Z would result in different $R_{value}$ "regions", independently of the object density or thickness; hence it would enable rough material differentiation. Table 4 shows $R_{value}$ for the different materials considered here, calculated from the simulations, and the theoretical ones, obtained from tabulated values [21]. The $R_{value}$ calculated from the simulations of both convertors are in good agreement with the theoretical ones, for the three $R_{value}$-regions examined (low ($R_{value}$ ~0.6), medium ($R_{value}$ ~0.9) and high Z ($R_{value}$ ~1.35)).

**Table 4: Theoretical and calculated $R_{Value}$. The error columns show the $\frac{R_{Value}^{Theory} - R_{Value}^{Simulation}}{R_{Value}^{Theory}}$.**

| Material | $R_{value}$ | | | | |
|---|---|---|---|---|---|
| | Theory | LXe volume simulation | Error | Tefzel capillaries simulation | Error |
| Pb | 1.34 | 1.37 | -1.9% | 1.37 | -1.9% |
| W | 1.33 | 1.32 | 0.4% | 1.33 | -0.2% |
| U | 1.35 | 1.37 | -1.5% | 1.33 | 1.4% |
| Polyethylene | 0.56 | 0.53 | 5.3% | 0.54 | 2.7% |
| Graphite | 0.59 | 0.55 | 5.5% | 0.55 | 6.1% |
| Al | 0.73 | 0.72 | 1.4% | 0.79 | -8.0% |
| PETN | 0.61 | 0.67 | -10.5% | 0.67 | -10.2% |
| Fe | 0.96 | 0.98 | -2.3% | 0.97 | -0.9% |
| Si | 0.75 | 0.75 | 0.0% | 0.72 | 4.1% |

**Fast-neutron** resonance radiography (FNRR [4]) exploits the differences in neutron's cross-sections with energy, of different elements, to identify specific elements within inspected items. For example, the neutron total cross-section for carbon has resonances in the energy range of 7.7-8.83MeV and dips in the range of 6.85-7.2MeV. A simplistic procedure of dividing the image received with neutrons of 6.85-7.2MeV by the one with neutrons of 7.7-8.83MeV would emphasize materials containing high carbon concentration, e.g. graphite (see figure 16-A for plain LXe converter and figure 17-A for Tefzel capillaries convertor). Similarly, the neutron cross-section for oxygen has resonances in the energy range of 3.26-3.79MeV and dip in the 2.31-2.37MeV range. Dividing the image recorded with neutrons of 2.31-2.37MeV by the one with neutrons of 3.26-3.79MeV would emphasize materials containing high oxygen concentration, e.g.

a PETN explosive in our case (see figure 16-B for plain LXe converter and figure 17-B for Tefzel capillaries convertor). The Tefzel capillaries convertor is similar to the plain LXe converter in terms of neutron's detection efficiency (see figure 13B) but permits better resolution ratio-images.

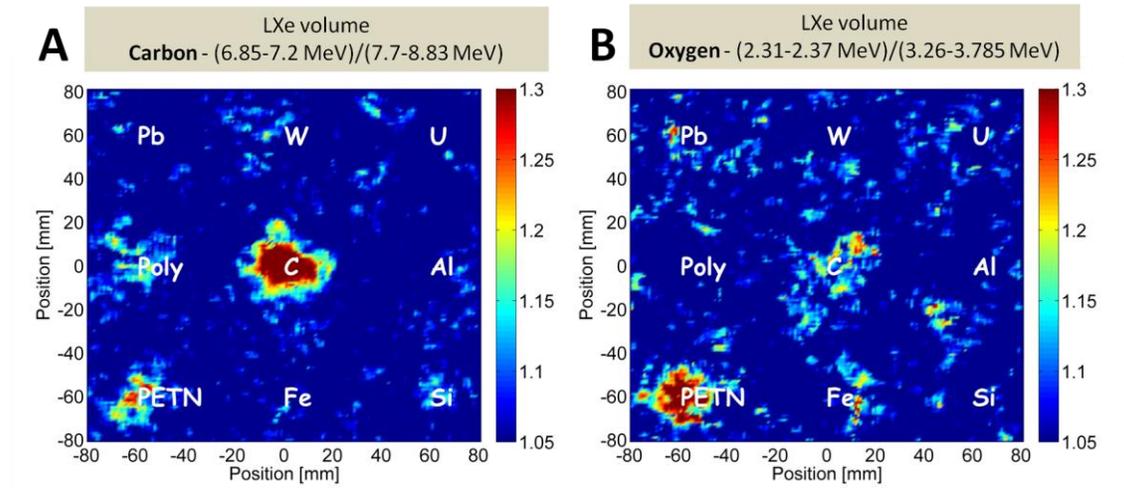

**Figure 16: Simulation results of a detector with plain LXe converter. Material differentiation using fast-neutron resonance radiography. (A) ratio between two images recorded with neutrons of 6.85-7.2MeV and 7.7-8.83MeV, emphasizing the graphite object. (B) ratio between two images recorded with neutrons of 2.31-2.37MeV and 3.26-3.79MeV emphasizing the oxygen-rich (explosive) object.**

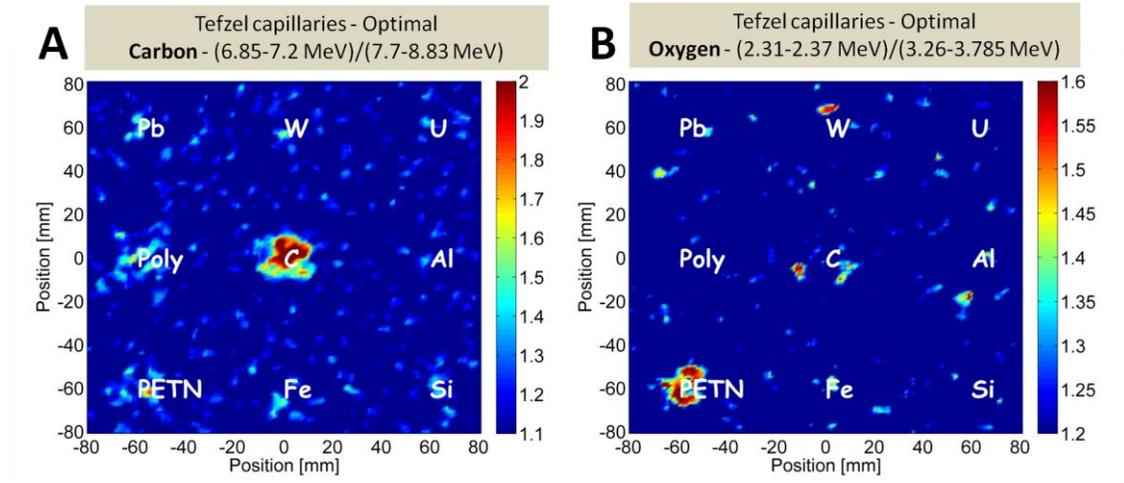

**Figure 17: Simulation results of a detector with Tefzel capillaries convertor of the optimal geometry. Material differentiation using fast-neutron resonance radiography. (A) ratio between two images recorded with neutrons of 6.85-7.2MeV and 7.7-8.83MeV, emphasizing the graphite object. (B) ratio between two images recorded with neutrons of 2.31-2.37MeV and 3.26-3.79MeV emphasizing the oxygen-rich (explosive) object.**

A much more elaborate approach, using several energies has been used successfully by the PTB/Soreq group [29] and will be tried using our simulation data.

**5. Discussion and Conclusions**

This work aimed at evaluating, through extended simulations, the expected performance of a combined fast-neutron and gamma-ray imaging detector concept for the 2-15 MeV energy range. It encompasses [9] an efficient fast liquid-xenon (LXe) converter-scintillator, coupled to a UV-sensitive gaseous imaging photomultiplier (GPM) with a CsI photocathode deposited on the top face of a cascaded THGEM multiplier. The main foreseen application is the simultaneous detection of concealed explosives (low-Z elements, with fast neutrons) and fissile materials (high-Z elements, with gammas).

Systematic GEANT4 simulations were carried out to evaluate detector parameters such as: deposited energy spectroscopy, detection efficiency and spatial resolution for 50 mm thick plain LXe and LXe-filled capillary converters.
Simulations of deposited energy spectra showed that:
- For gamma-rays, only the plain LXe converter exhibits peaks (full photo-peak, single and double escape) in the spectrum. The introduction of capillaries causes significant modification in the spectrum, since the gamma-ray induced electrons/positrons created in LXe deposit only part of their energy within the scintillator volume. Thus, the pulse height distribution expected with a capillary converter will not show peaks. In case of DDEGR, where we use only two energies (4.43 MeV and 15.1 MeV), this should not present a problem since the two gamma-ray energies are sufficiently distant from each other, as demonstrated in figure 14.
- Deposited energy neutron spectra are continuous in all converters. The addition of hydrogenous media in the form of Tefzel or polyethylene capillaries extends the spectra to larger deposited energy values. As we intend to perform neutron spectroscopy by TOF, the pulse height analysis of neutrons is of less relevance here.

The simulations of detection efficiency indicated that:
- Detection efficiencies of gammas in Teflon, Tefzel and polyethylene converters are ~35% . The plain-LXe converter provided the highest gamma-ray detection efficiency, of ~40-50% for 2-15MeV energy range.

- Neutron detection efficiency for Teflon, Tefzel and plain LXe converters is about 20% over the entire energy range. In hydrogen-rich polyethylene capillaries the neutron detection efficiency for low neutrons energy (2 MeV) was lower by a factor of ~2. Our conclusion is that the reduction of efficiency at low neutron energy is caused by the lower average deposited energy available for production of light (see table 1).

Calculations of detectors' spatial resolution showed that:
- For gamma radiation similar spatial resolutions were obtained in all four different detector configurations (FWHM of 2-5mm) due to the small cross sections for gamma interaction in Teflon, Tefzel and Polyethylene. The optimal, commercially-available, Tefzel-capillaries (inner radius of 0.508 mm and outer radius of 0.794 mm) resulted in spatial resolution of ~2.4mm (FWHM) at 4.4MeV and ~3.5mm (FWHM) at 15.1MeV.
- For neutrons, the narrowest spatial distribution was obtained with the Polyethylene or Tefzel capillaries (FWHM of ~2.5mm), due to higher energy transfer to the capillary materials close to the point of interaction and higher photons statistics in case of scintillation from knock-off-protons. The optimal, commercially-available, Tefzel-capillaries with the above dimensions resulted in spatial resolutions of ~1.5mm (FWHM), for 2-10MeV neutrons, which is only marginally inferior compared to that obtained with TRION detector (~1 mm FWHM).
  The broadest spatial resolution was (~8mm FWHM) obtained with plain LXe converter.

Both detection efficiency and the spatial resolution obtained here are adequate for the intended application of explosives and SNM detection.

An initial attempt was made to evaluate the potential of elemental discrimination by simulating radiographic images of various materials using two discrete energy gamma-rays (4.4 MeV and 15.1 MeV) and neutrons in broad energy range (2-10 MeV). The results indicated that:

- Dual-energy gamma radiography in both detectors, with plain LXe volume and Tefzel capillaries radiation-converters, resulted in $R_{value}$ which are in good (0-10%) agreement with those derived from the tabulated cross-sections values, for the three $R_{value}$-regions examined (low ($R_{value}$ ~0.6), medium ($R_{value}$ ~0.9) and high ($R_{value}$ ~1.35) Z numbers),

- Fast-neutron radiographic images, simulated at different neutron energies, chosen to cover the resonance cross-section features of low-Z elements, indicated the potential of elemental discrimination. A simple procedure of utilizing image division ON and OFF resonance energy demonstrated the detector's ability of identifying elements such as C and O. A much more elaborate multi-energy reconstruction approach will be evaluated in the future.

An experimental system (WILiX [30]) based on a cryostat incorporating a LXe converter coupled to a 4" diameter GPM has been completed and it is fully operational. It will permit validating the detector concept and to carry out imaging experiments with fast neutrons and gammas at the laboratory and with accelerator beams.

**Acknowledgments**

This work was partly supported by the Minerva Foundation with funding from the German Ministry for Education and Research (Grant No. 710827), the Israel Science Foundation (Grant No. 477/10) and the PAZY Foundation (Grant No. 258/14). A. Breskin is the W.P. Reuther Professor of Research in The Peaceful Use of Atomic Energy.